# Pressure-induced superconductivity and topological phase transitions in the topological nodal-line semimetal SrAs$_3$

E. J. Cheng[1,*], W. Xia[2,4,10*], X. B. Shi[3,5*], Z. H. Yu[2], L. Wang[6], L. M. Yan[6], D. C. Peets[7], C. C. Zhu[1], H. Su[2], Y. Zhang[8], D. Z. Dai[1], X. Wang[9], Z. Q. Zou[9], N. Yu[9], X. F. Kou[8], W. G. Yang[6], W. W. Zhao[3,5,†], Y. F. Guo[2,†] and S. Y. Li[1,11†]

[1] *State Key Laboratory of Surface Physics, Department of Physics, and Laboratory of Advanced Materials, Fudan University, Shanghai 200438, China*

[2] *School of Physical Science and Technology, ShanghaiTech University, Shanghai 200031, China*

[3] *State Key Laboratory of Advanced Welding & Joining and Flexible Printed Electronics Technology Center, Harbin Institute of Technology, Shenzhen 518055, China*

[4] *Shanghai Institute of Optics and Fine Mechanics, Chinese Academy of Sciences, Shanghai 201800, China*

[5] *Key Laboratory of Micro-systems and Micro-structures Manufacturing of Ministry of Education, Harbin Institute of Technology, Harbin 150001, China*

[6] *Center for High Pressure Science and Technology Advanced Research, Shanghai 201203, China*

[7] *Ningbo Institute of Materials Technology and Engineering, Chinese Academy of Sciences, Ningbo, Zhejiang 315201, China*

[8] *School of Information Science and Technology, ShanghaiTech University, Shanghai 201210, China*

[9] *Analytical Instrumentation Center, School of Physical Science and Technology, ShanghaiTech University, Shanghai 201210, China*

[10] *University of Chinese Academy of Sciences, Beijing 100049, China*

[11] *Collaborative Innovation Center of Advanced Microstructures, Nanjing 210093, China*

[*] *These authors contributed equally to this work.*

**Topological nodal-line semimetals (TNLSMs) are materials whose conduction and valence bands cross each other, meeting a topologically-protected closed loop rather than discrete points in the Brillouin zone (BZ). The anticipated properties for TNLSMs include drumhead-like nearly**

**flat surface states, unique Landau energy levels, special collective modes, long-range Coulomb interactions, or the possibility of realizing high-temperature superconductivity. Recently, SrAs$_3$ has been theoretically proposed and then experimentally confirmed to be a TNLSM. Here, we report high-pressure experiments on SrAs$_3$, identifying a Lifshitz transition below 1 GPa and a superconducting transition accompanied by a structural phase transition above 20 GPa. A topological crystalline insulator (TCI) state is revealed by means of density functional theory (DFT) calculations on the emergent high-pressure phase. As the counterpart of topological insulators, TCIs possess metallic boundary states protected by crystal symmetry, rather than time reversal. In consideration of topological surface states (TSSs) and helical spin texture observed in the high-pressure state of SrAs$_3$, the superconducting state may be induced in the surface states, and is most likely topologically nontrivial, making pressurized SrAs$_3$ a strong candidate for topological superconductor.**

In recent years, topological semimetals including Dirac, Weyl, and nodal-line semimetals have been theoretically predicted and experimentally verified, opening a new field in condensed-matter physics in which novel properties and new applications can arise from spin-polarized states with unique band dispersion[1-3]. Unlike the discrete points in momentum space in Dirac or Weyl semimetals[2,3], the band crossings in nodal-line semimetals can form closed loops inside the BZ[4]; a nodal chain consisting of several connected loops[5]; or an extended line traversing the entire BZ[6]. These one-dimensional nodal curves are topologically protected by certain discrete symmetries, for example mirror reflection, time-reversal, or spin-rotation symmetries[2,3]. Upon breaking symmetries in a TNLSM, the nodal line is either fully gapped or gapped into several nodal points[4]. The nodal-line structure is expected to have several intriguing properties[3], such as unique Landau energy levels[7], special collective modes[8], long-range Coulomb interactions[9], or drumhead-like nearly-flat surface states[10,11] which can be considered a higher-dimensional analogue of the flat band on the zigzag edge of graphene[3]. These drumhead states may host interesting correlation effects, and even offer the possibility of realizing high-temperature superconductivity[12].

In the search for nodal-line semimetals, several systems have been theoretically proposed since 2011[2,3]. However, only a few candidates including PbTaSe$_2$[13,14], ZrSi$X$ ($X$ = S, Se, Te)[15,16], CaAg$X$ ($X$ = P, As)[17,18] and $M$B$_2$ ($M$ = Ti, Zr)[19] have been verified experimentally. More recently, the CaP$_3$ family of materials ($M$As$_3$, for $M$ = Ca, Ba, and Sr$X_3$ for $X$ = P, As) was proposed as another potential host of TNLSMs[20]. Among these compounds, only SrAs$_3$ shows a strongly topological nature at ambient pressure, while others need extra compressing[20]. SrAs$_3$ displays semimetallic behavior with the hole carriers dominating[21,22]. Previously, unusual galvanomagnetic properties and a first-order longitudinal Hall effect have been found in SrAs$_3$[23], and quantum oscillation experiments have been applied to map out the shape of the Fermi surface, finding two asymmetric, quasi-ellipsoidal Fermi-bodies as well as light cyclotron effective mass[22]. Recent magnetotransport measurements on SrAs$_3$ single crystals found a nontrivial Berry phase and a robust negative longitudinal magnetoresistance (MR) induced by the chiral anomaly, which indicates the presence of topological properties in SrAs$_3$[24,25]. Subsequently, Song *et al.* observed the complete nodal-line feature around the $Y$ point by means of angle-resolved photoemission spectroscopy (ARPES), demonstrating the existence of Dirac nodal-line fermions[26]. In contrast to most TNLSMs, the nodal-line structure in SrAs$_3$ does not coexist with complex topologically trivial Fermi surfaces, which may pave an easy path to potential applications[20,26].

Among the topological materials, intense effort has been applied to realizing topological superconductors (TSCs)[27,28], one source of Majorana fermions, an effort which suffers from a severe lack of suitable materials to study[27,28]. Experimentally, applying chemical doping or pressure to search for superconductivity in known topological materials are two common methods to obtain new TSC candidates[27,28]. While chemical doping introduces chemical complexity and disorder, pressure is a clean and effective approach for tuning the interactions among multiple degrees of freedom, and superconductivity has been found in many materials via this route[29-31]. Among the CaP$_3$ family of materials, CaAs$_3$ was proposed to host a single nodal loop due to time reversal, spatial inversion, and accidental degeneracies[32]. Li *et al.* reported its transport properties under hydrostatic pressure up to 2.09 GPa, finding a decrease in the resistivity and a possible superconducting transition under pressure[33]. Since SrAs$_3$ has already been demonstrated to be a TNLSM, the lack of any high-pressure report inspired us to explore its pressure dependence.

In this work, we present the results of high-pressure measurements on single-crystalline SrAs$_3$. Upon applying pressure, the topologically-protected $\alpha$ pocket and trivial $\beta$ pocket disappear around 1 GPa, and two higher frequencies denoted as $\varepsilon$ and $\zeta$ emerge, indicating a Lifshitz transition. More interestingly, a superconducting transition has been observed from 20.6 GPa, with a dome-like pressure dependence. High-pressure X-ray diffraction (XRD) was conducted to investigate the high-pressure structure of SrAs$_3$, and a structural transition was found around 20 GPa. DFT calculations on the high-pressure structure of SrAs$_3$ reveal a TCI state. TCI states have previously been experimentally verified only in narrow-gap IV-VI semiconductors with a rock-salt structure, for example SnTe[34] and Pb$_{1-x}$Sn$_x$M (M = Se, Te)[35,36]. Thus, the observation of a pressure-induced TCI state in SrAs$_3$ offers an alternative route to explore this exotic state. Moreover, the finding of superconductivity in this state makes high-pressure SrAs$_3$ a candidate topological superconductor.

## Results

SrAs$_3$ crystallizes in a triclinic (space group $P\bar{1}$) or monoclinic (space group *C2/m*) structure; the latter is proposed to possess topological-nodal-line states protected by time-reversal symmetry, spatial-inversion symmetry, and mirror symmetry[20]. Figure 1(a) shows the unit cell of monoclinic SrAs$_3$. This crystal structure can be viewed as a stack of two-dimensional (2D) infinite polyanionic layers $^2_\infty[P_3]^{2-}$ along the *b* axis[21]. The As layers form channels and the Sr cations are inserted into the channels, as shown in Fig. 1(b)[21]. The inset in Fig. 1(c) shows the x-ray diffraction (XRD) rocking curves of SrAs$_3$ single crystals grown from both Bi flux (which we refer to as BF) and self flux (SF). From an x-ray rocking curve of the (002) Bragg peak, a full width at half maximum (FWHM) of 0.04 ° indicates the high quality of the SrAs$_3$ single crystal grown from Bi flux, while a broader FWHM of 0.15 ° for the SF sample suggests lower quality. In resistivity measurements (Fig. 1(c)), the SF sample exhibits semimetallic behavior, while the BF sample exhibits metallic behavior with a residual resistivity ratio RRR = $\rho(300\,\text{K})/\rho_0$ of 5, and a residual resistivity $\rho_0$ of 72.7 μΩ cm, which are further indicative of its high quality. For magnetotransport measurements, the higher-quality BF-grown samples were chosen.

Figure 2(a) shows the temperature-dependent resistivity from 1.8 to 300 K at pressures up to 1.47 GPa on a BF-grown SrAs$_3$ single crystal. Upon increasing the pressure from 0.14 to 1.18 GPa, the low-temperature resistivity becomes increasingly metallic. The MR, shown in Fig. 2(b), is non-monotonic in pressure. Upon increasing the pressure from 0.14 to 1.18 GPa, the MR at 9 T and 1.8 K increases from 5000% to 32500%. However, further increasing the pressure to 1.47 GPa reduces the MR to 22000%. The oscillatory component $\Delta R_{xx}$, plotted in Fig. 2(c), exhibits low-frequency modes for lower pressures and higher-frequency modes at higher pressures, indicating a significant change in the electronic structure at the Fermi surface. Fast Fourier transforms (FFTs) of the MR oscillations, displayed in Fig. 2(d) contain only the *α* (1.4 T) and *β* (5.5 T) pockets from 0.14 to 0.69 GPa, as previously seen at ambient pressure[25]. At 0.99 GPa, the $\alpha$ and $\beta$ pockets abruptly disappear, replaced by a single frequency of 21.5 T which we assign to a $\zeta$ pocket. Upon increasing the pressure to 1.47 GPa, the $\zeta$-frequency pocket is joined by an even higher frequency of 48.3 T which grows rapidly to 63.2 T, which we assign to an $\varepsilon$ pocket.

Figure 2(e) summarizes the pressure-dependent oscillation frequencies, with shading identifying the transition where the Fermi surface changes. The two obvious scenarios for this abrupt transition are a structural transition or a Lifshitz transition. But high-pressure diffraction, shown in Fig. 4 and discussed in more detail below, indicates that the ambient-pressure structure persists to ~20 GPa, excluding a structural origin. In topological materials, different types of Lifshitz transitions are possible, involving other types of zeroes in the energy spectrum in addition to or instead of the Fermi surface, such as flat bands, Weyl and Dirac nodes, Dirac nodal lines, zeroes in the spectrum of edge states, Majorana modes, etc[37]. For example, in rhombohedral graphite, the Lifshitz transition is related to the change of the shape of the Dirac line[37]. To study the relationship between the Lifshitz transition and the topologically-protected nodal-line structure in SrAs$_3$, a Landau level index fan diagram for the pressure-induced pockets is plotted in Fig. 2(f), where no Berry phase has been observed for either the $\varepsilon$ or $\zeta$ pocket. To produce this diagram, we assign integer indices to the $\Delta R_{xx}$ peak positions in $1/B$ and half-integer indices to the $\Delta R_{xx}$ valleys. Infinite-field intercepts around 3/8 to 5/8 indicate trivial band topology. In SrAs$_3$, the $\alpha$ pocket is topologically protected, and the nodal line is situated around the $Y$ point in the BZ[24]. Hence, we surmise that the Fermi surface of SrAs$_3$ is reconstructed below 1.0 GPa, in a Lifshitz transition which involves the topologically-protected nodal-line structure.

As a TNLSM, SrAs$_3$ possesses 2D drumhead-like nearly-flat surface states nested inside the closed nodal line, and it may provide an ideal playground for many interaction-induced nontrivial states, such as superconductivity. In the resistivity under higher pressures in Fig. 2(a), there is a clear suppression of the low-temperature resistivity. Pressure-induced superconductivity in topological semimetals often appears as the extremely large MR is suppressed[31]. Given the significant reduction in MR from 1.18 to 1.47 GPa, we suspected that higher pressures may induce superconductivity. Since bismuth superconducts under pressure, with $T_c$s of 3.92 and 7 K for pressures of 2.63 and 2.8 GPa, respectively[38], self-flux-grown SrAs$_3$ crystals were studied to completely eliminate the possibility of Bi flux inclusions. Figure 3(a) shows the resistance curves for SrAs$_3$ under higher pressures. No superconducting transition is observed till 20.6 GPa. At 20.6 GPa, a superconducting transition with

$T_c^{10\%}$ of 3.6 K appears, and the $T_c$ increases to 5.8 K at 54.7 GPa. Upon further pressurization to 63.6 GPa, the $T_c$ decreases slightly to 5.5 K. The pressure dependence of the superconducting transition, summarized in Fig. 3(c), is clearly dome-shaped. To verify that this is a superconducting transition, the effect of magnetic field at 39.4 GPa was studied, as plotted in Fig. 3(b) – the transition is gradually suppressed by magnetic field, as expected for superconductivity. Figure 3(d) plots the temperature dependence of the upper critical field ($\mu_0 H_{c2}$). We used the Ginzburg-Landau (GL) formula $\mu_0 H_{c2}(T) = \mu_0 H_{c2}(0)(1-(T/T_c)^2)/(1+(T/T_c)^2)$ to fit the data at 39.4 GPa; the $\mu_0 H_{c2}(0)$ values are estimated to be 2.58(2), 2.00(2), and 1.48(3) T for $T_c^{10\%}$, $T_c^{50\%}$, and $T_c^{zero}$, respectively, yielding coherence length $\xi_{GL}(0)$ of 11.3, 12.9, and 14.9 nm. These fields are much lower than the Pauli limiting fields[39,40] $H_P(0) = 1.84T_c \sim 9.5$, 7.9, and 6.9 T, respectively, indicating that Pauli pair breaking is not relevant.

To verify whether the pressure-induced superconductivity arises from a structural phase transition, we performed high-pressure synchrotron XRD measurements, shown in Fig. 4(a) for different pressures. The ambient-pressure (denoted as AP) phase persisted up to ~ 20 GPa. New diffraction peaks (marked with a dashed line and asterisk) appear at ~ 23.3 GPa, indicating a pressure-induced structural phase transition, and these continue to strengthen upon further increasing pressure. Figure 4(b) depicts the crystal structure of SrAs$_3$ under ambient conditions. There are two Wyckoff positions (4i and 8j) for As atoms in each unit cell, and Sr (4i) atoms are coordinated to seven As atoms. The crystal structure of the ambient-pressure phase below 20.4 GPa was refined according to the initial model determined by room temperature single-crystal X-ray diffraction. Typical GSAS refinements for SrAs$_3$ under 3.6 GPa are illustrated in Fig. 4(c). In general, applying pressure tends to increase the coordination number (CN) of cations, the so-called "pressure-coordination" rule, as seen for instance in Bi$_2$Se$_3$, where *R-3m* (CN = 6) transforms to *C2/m* (CN = 7) and eventually *I4/mmm* (CN=8)[41] with pressure. A further effect, the "corresponding static principle", states that compounds containing light elements tend at higher pressures to adopt the structures observed at lower pressures in analogous compounds with heavy elements of the same group. A typical example is found in the binary antimonides Li$_3$X (X = N, P, As, Sb and Bi): Li$_3$N crystallizes in the hexagonal *P6/mmm* structure at ambient pressure and transforms into the hexagonal *P6$_3$/mmc* structure (which is the ambient-pressure structure of Li$_3$P, Li$_3$As and Li$_3$Sb) around 0.5 GPa. It then undergoes a second structural phase transition to a cubic *Fm–3m* structure (the structure of Li$_3$Bi at ambient pressure) around 36 ~ 45 GPa[42].

At ambient conditions, SrP$_3$ and SrAs$_3$ have the same crystal structure (*C2/m*), while SrBi$_3$ forms in the cubic Cu$_3$Au-type structure (space group: $Pm$-$3m$, No. 221)[43]. Thus the corresponding static principle suggests the Cu$_3$Au structure type as a strong candidate for the high-pressure structure. The crystal structure model for high-pressure SrAs$_3$ was deduced by testing several candidates, and was ultimately refined with a SrBi$_3$-like structure (space group: $Pm$-$3m$, No. 221)[43]. The schematic crystal structure of the high-pressure phase is depicted in Fig. 4(d). Sr and As atoms occupy 1a (0, 0, 0) and 3c (0, 0.5, 0.5) Wyckoff positions, respectively, and the CN for Sr has increased to 12. Figure 4(e) shows the Rietveld refinement of SrAs$_3$ under 49.8 GPa, yielding 70% and 30% for the ambient and high-pressure phases, respectively. The coexistence of high and ambient-pressure phases up to 51.9 GPa indicates that the pressure-induced structural phase transition in SrAs$_3$ is first order and that the two phases have only a minute difference in Gibbs free energy.

Since the pressure-induced superconductivity appears long after the topologically-protected $\alpha$ pocket is eliminated, it is important to check whether the high-pressure band structure of SrAs$_3$ is topologically nontrivial, which is a necessary condition for topological superconductivity. To obtain more electronic structure information on the high-pressure phase of SrAs$_3$, we performed DFT calculations for pressurized SrAs$_3$ at 34 GPa, as summarized in Fig. 5. The band structure in the absence of spin-orbit coupling (SOC), shown in Fig. 5(a), displays metallic behaviour. The electron-dominant valence and hole-dominant conduction bands cross near the Fermi level along the $\Gamma$-$M$ line. Owing to the cubic symmetry of this material, there are twelve band crossing points at symmetrically-equivalent points in the full BZ. Upon turning on SOC, a gap will be opened at these crossing points, resulting in a continuous SOC gap with a curved chemical potential between valence and conduction bands at each $k$ point, as shown in Fig. 5(b).

To identify the topological nature of this material, the Fu-Kane parity criterion[44,45] at eight time-reversal invariant momenta (TRIM) was utilized to determine the $Z_2$ index. We obtain a trivial $Z_2$ of (0;000) from the production of the parities of all occupied bands at the eight TRIM points, as shown in Fig. 5(c). However, the surface states on the (001)-projected surface contain two surface Dirac cones located at $\bar{X}$ points, as seen in Fig. 5(d). Because of SOC, these Dirac states host a helical spin texture as shown in Fig. 5(f). The Wilson loop method was employed to determine the mirror Chern number (MCN), and get MCN = 1, in agreement with the surface state behavior observed on the (001) surface[44-46]. The continuous SOC gap, topologically-trivial $Z_2$ index, nontrivial mirror Chern number and even number of surface Dirac points indicate that this material is a topological crystalline insulator[44-46]. As a counterpart of topological insulators in which crystalline symmetry replaces time-reversal symmetry to enforce topological protection, TCIs possess topological surface states (TSSs) with an even number of gapless Dirac cones on the surface BZ and host a variety of exotic phenomena[46], for example, large-Chern-number quantum anomalous Hall effect[47] or strain-induced superconductivity[48]. The superconductivity in high-pressure SrAs$_3$ may exist in or be induced in these surface states. Owing to their helical spin texture, any superconducting phase in these surface states would most likely be topologically nontrivial[34-36,46-48], making high-pressure SrAs$_3$ a strong candidate for topological superconductor.

## Discussion

Searching for Majorana fermions has been fueled by the prospect of using their non-Abelian statistics for robust quantum computation, and they can be realized as a bound state at zero energy, i.e. Majorana bound states (MZM), in the vortex core of a topological superconductor[27,28]. To realize a topological superconductor, two primary routes have been proposed, i.e. bulk spin-triplet superconductivity, and superconductivity in spin-nondegenerate TSSs induced by the proximity effect, for instance through heterostructures stacking conventional *s*-wave superconductors and topological insulators, quantum anomalous Hall insulators, nanowires, or atomic chains[27,28,49]. In the former case, due to the breaking of spin degeneracy by asymmetric SOC, the parity-mixed superconducting state in non-centrosymmetric superconductors may also host Majorana fermions if the singlet component is smaller than the triplet component[50]. Recently, the layered non-centrosymmetric compound PbTaSe$_2$ with strong SOC was reported to possess fully-gapped multiband superconductivity[51]. However, its spin-triplet component is small or absent[52,53]. Bulk topological nodal-line states and fully spin-polarized TSSs have been also demonstrated in PbTaSe$_2$[13], which may allow proximity-induced fully-gapped superconducting TSSs, which could pair in $p_x + ip_y$ symmetry and host bound Majorana fermions in the vortices[14].

The superconducting phase of SrAs$_3$ is centrosymmetric (*Pm-3m*), so barring the exceedingly unlikely possibility of bulk spin-triplet superconductivity, the bulk route to Majorana quasiparticles is not available. However, we propose a TCI state hosting TSSs with helical spin texture. Superconductivity in these states or induced in these states by proximity effect from the bulk could potentially be topologically nontrivial. However, unlike in other TCIs such as SnTe or Pb$_{1-x}$Sn$_x$*M* (*M*

= Se, Te), trivial bulk bands in pressurized $SrAs_3$ cross the Fermi level, and we are unable to distinguish which bands participate in the superconductivity. Further work will be required to elaborate the contributions from TSSs and bulk states to the superconductivity.

In summary, at ambient pressure, $SrAs_3$ is a TNLSM with 2D drumhead-like nearly-flat surface states, which may be strongly correlated and are often associated with the enhancement of superconductivity[54], although pressure did not succeed in driving this material superconducting before changing the electronic structure. A Lifshitz transition has been identified below 1.0 GPa, evidencing a topological phase transition, as the quantum oscillations associated with the TNLSM state vanish. Higher-pressure experiments on a powder sample reveal a dome-like superconducting transition accompanying a structural phase transition into a phase which we predict to host a topological-crystalline-insulator state with TSSs and helical spin texture. Besides its intrinsic interest as a TNLSM, $SrAs_3$ offers an alternative route to explore the topological-crystalline-insulator state beyond IV-VI semiconductors and, as a superconducting TCI, high-pressure $SrAs_3$ could serve as a candidate topological superconductor. Doping studies or strain may still be able to induce superconductivity in the low-pressure phase, and should be pursued, and the evolution of the drumhead-like states with pressure remains to be clarified.

## Methods

**Sample synthesis.**

**Self-flux method.** Sr (99.95 %, Alfa Aesar), and As (99.999 %, PrMat) were mixed in a molar ratio of 1:3 and placed into an alumina crucible. The crucible was sealed in a quartz ampoule under vacuum and subsequently heated to 750 °C in 10 h. After reaction at this temperature for 300 h, the ampoule was cooled to 400 °C in 50 h and cooled freely to room temperature. $SrAs_3$ single crystals with black shiny metallic lustre were obtained.

**Bi-flux method.** Sr (99.95%, Alfa Aesar), As (99.999%, PrMat) and Bi (99.9999%, Aladdin) blocks were mixed in a molar ratio of 1:3:26 and placed into an alumina crucible. The crucible was sealed in a quartz ampoule under vacuum and subsequently heated to 900 °C in 15 h. After reaction at this temperature for 20 h, the ampoule was cooled to 700 °C over 20 h, and then slowly cooled to 450°C at 1 °C/h. The excess Bi flux was then removed in a centrifuge, and $SrAs_3$ single crystals with black shiny metallic lustre were obtained.

**Pressure measurements.**

**Resistance measurements under pressure.** For high-pressure experiments, samples were loaded in a piston-cylinder clamp cell made of Be-Cu alloy, with Daphne oil as the pressure medium. The pressure inside the cell was determined from the $T_c$ of a tin wire. A SrAs$_3$ single crystal was cut into a bar shape, and the standard four-probe method was used for resistivity measurements, with contacts made using silver epoxy. Higher-pressure measurements were performed on powder samples comprising crushed single crystals using a diamond anvil cell (DAC). The experimental pressures were determined by the pressure-induced fluorescence shift of ruby[55] at room temperature before and after each experiment. A direct-current van der Pauw technique was adopted. Resistance measurements were performed with a physical property measurement system (PPMS, Quantum Design).

**X-ray diffraction (XRD) measurements under pressure.** SrAs$_3$ single crystals grown by the self-flux method were ground into fine powder in a mortar for use in the high-pressure synchrotron angle dispersive x-ray diffraction (AD-XRD) measurement. The high-pressure synchrotron XRD experiments were carried out using a symmetric diamond anvil cell (DAC) with a 260-micron culet diamond. A rhenium gasket was precompressed to 30 microns in thickness followed by drilling the central part by laser to form a 90-micron diameter hole as the sample chamber. The sample chamber was filled with a mixture of the sample, a ruby chip, and silicone oil as the pressure transmitting medium. The experimental pressures were determined by the pressure-induced fluorescence shift of ruby[55]. Synchrotron AD-XRD measurements were carried out at beamline BL15U1 of the Shanghai Synchrotron Radiation Facility (SSRF) using a monochromatic beam of 0.6199 Å.

**Density functional theory (DFT) calculations.** Density functional theory (DFT) calculations were performed based on the Perdew-Burke-Ernzerhof (PBE)-type generalized gradient approximation (GGA)[56], and used the projector augmented wave (PAW) method[57], as encoded in the Vienna *ab inito* simulation package (VASP)[58]. The cutoff energy for the plane-wave basis taken was 500 eV. The first Brillouin Zone was sampled, using a $\Gamma$-centered 12×12×12 $k$-point mesh. The energy convergence criteria were defined as 10$^{-8}$ eV. The lattice constants were fully relaxed using a conjugate gradient scheme until the Hellmann-Feynman forces on the ions were less than 0.001 eV/Å. We constructed the maximally-localized Wannier functions (MLWF)[59-61] using Sr $d$ and As $s$ and $p$ atomic orbitals. The topological features of surface state spectra were calculated using the iterative Green's function technique[62], as implemented in the Wannier-Tools package[63].

**Acknowledgements:**

This work is supported by the Ministry of Science and Technology of China (Grants No. 2015CB921401 and 2016YFA0300503), the Natural Science Foundation of China (Grant No. 11421404, 11674367, 11874264), the NSAF (Grant No. U1630248), the Natural Science Foundation of Shanghai (Grant No. 17ZR1443300), and the Zhejiang Provincial Natural Science Foundation (Grant No. LZ18A040002). W.W.Z. is supported by the Shenzhen Peacock

Team Plan (KQTD20170809110344233) and Bureau of Industry and Information Technology of Shenzhen through the Graphene Manufacturing Innovation Center (201901161514). D.C.P. is supported by the Chinese Academy of Sciences through 2018PM0036. Y.F.G. acknowledges the research fund from the State Key Laboratory of Surface Physics and Department of Physics, Fudan University (Grant No. KF2019_06). Y.F.G. and X.F.K. are also funded by the strategic Priority Research Program of Chinese Academy of Sciences (Grant No. XDA18000000).

**Author Contributions:** S.Y.L. and Y.F.G. conceived the idea and designed the experiments. E.J.C. was responsible for high-pressure transport experiments. W.X., Y.Z., X.W., Z.Q.Z., H.S. and X.F.K. performed sample synthesis and partial data analysis. Z.H.Y., N.Y., L.W., L.M.Y. and W.G.Y. conducted the high-pressure X-ray diffraction measurements and structure analysis. C.C.Z. and D.Z.D. helped the data collection. S.Y.L., Y.F.G. and W.W.Z. supervised the project. E.J.C., D.C.P., Y.F.G. and S.Y.L. analyzed the data and wrote the paper. E.J.C., W.X., and X.B.S. contributed equally to this work. All authors discussed the results and commented on the manuscript.

**Additional Information:** Correspondence and requests for materials should be addressed to W. W. Zhao (wzhao@hit.edu.cn), Y. F. Guo (guoyf@shanghaitech.edu.cn) and S. Y. Li (shiyan_li@fudan.edu.cn).

**Competing financial interests:** The authors declare no competing financial interests.

Figure 1

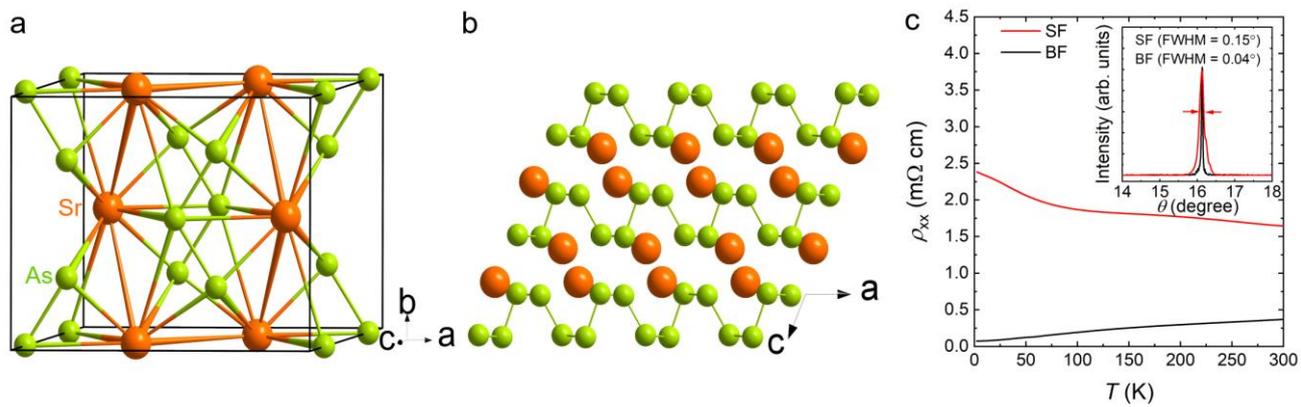

Figure 2

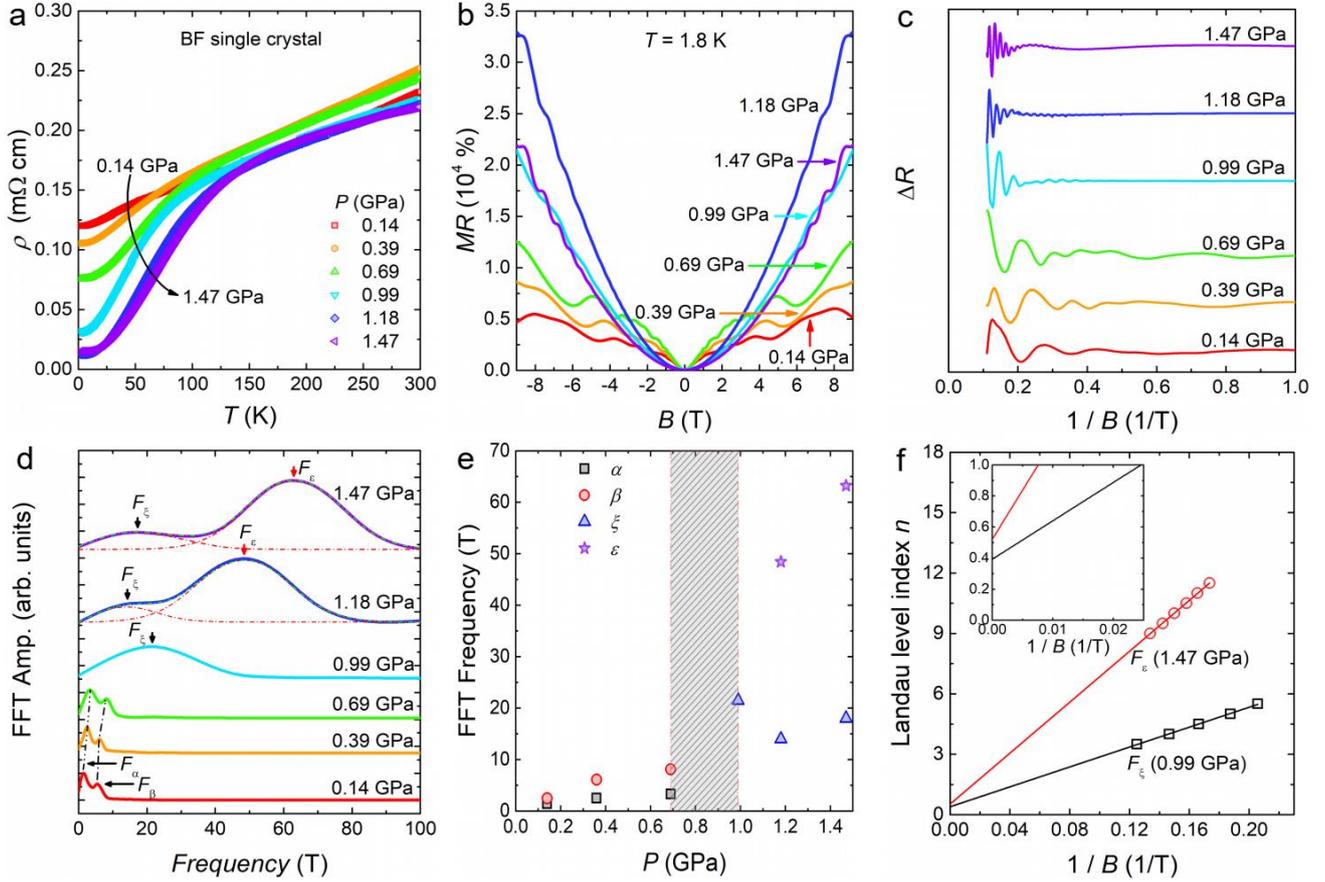

Figure 3

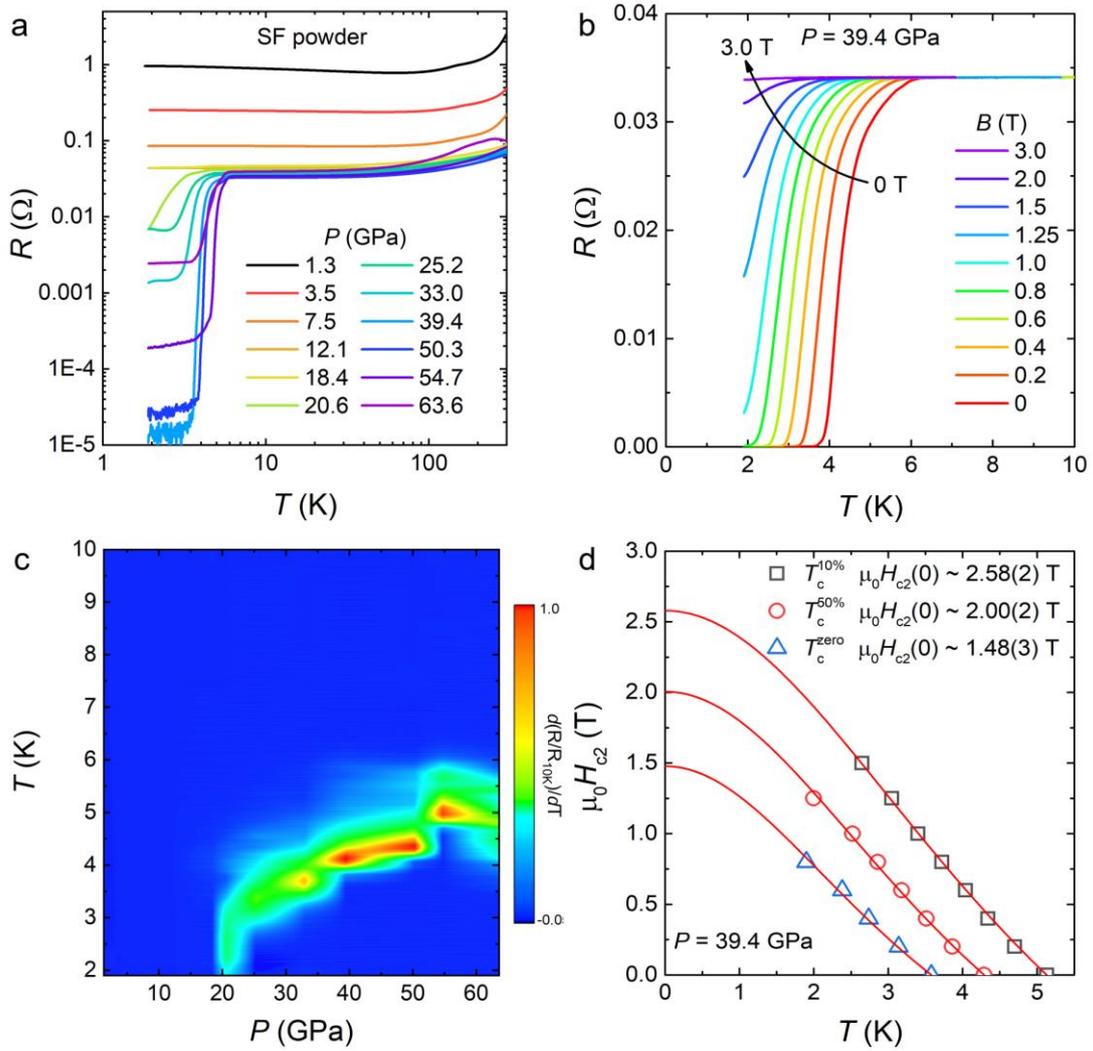

Figure 4

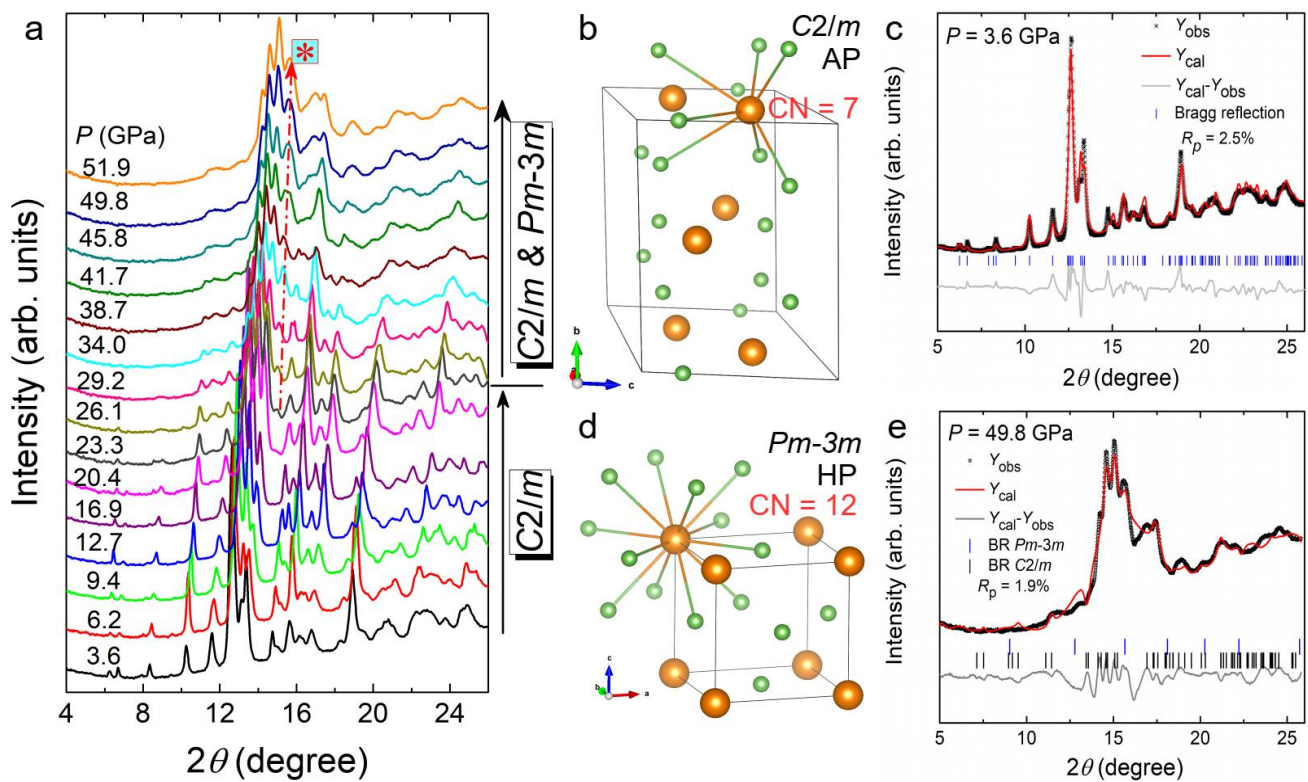

Figure 5

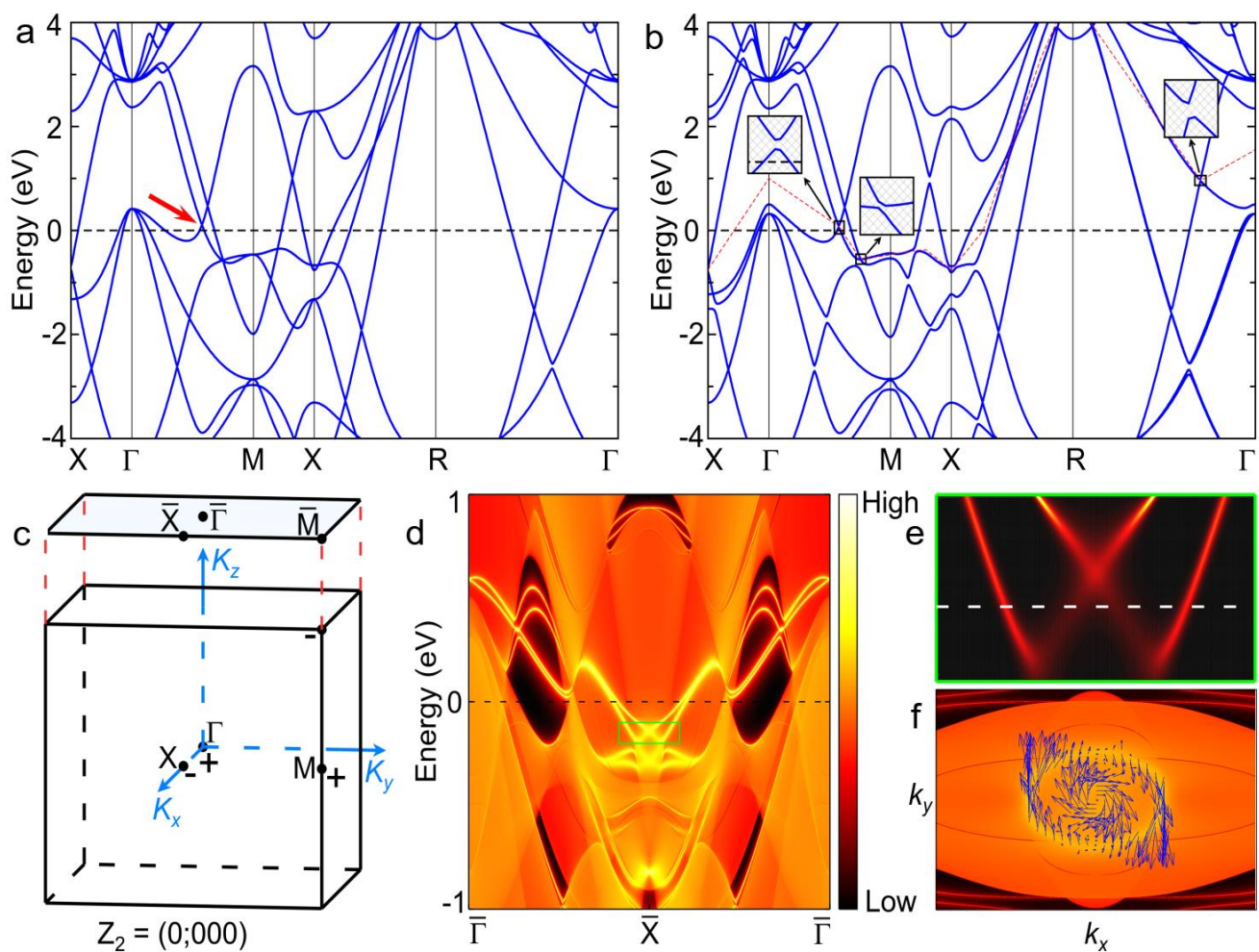

**Figure captions**

**Fig. 1 | Characterization of SrAs₃.** (**a**) The unit cell of monoclinic $SrAs_3$. (**b**) The side view along the *b* axis. (**c**) Longitudinal resistivity of $SrAs_3$ single crystal. The black and red curves represent data collected on $SrAs_3$ single crystals obtained through Bi (BF) and self-flux (SF) methods, respectively. The red curve (SF) shows semimetallic behaviour, while the other one (BF) displays metallic behavior. Inset: X-ray diffraction (XRD) rocking curves of the (002) Bragg peak for Bi-flux and self-flux single crystals. The full width at half maximum (FWHM) of Bi-flux single crystal is 0.04°, while the FWHM of self-flux single crystal is 0.15°, indicating that the quality of Bi-flux single crystals is higher.

**Fig. 2 | Shubnikov–de Haas (SdH) oscillation study of SrAs₃ under low pressure.** (**a**) Resistivity of a $SrAs_3$ single crystal grown from Bi-flux. Upon increasing the pressure to 1.18 GPa, the low-temperature resistivity is monotonously reduced. At 1.47 GPa, the resistivity slightly increases, relative to that at 1.18 GPa. (**b**) Pressure dependence of magnetoresistance (MR) of a $SrAs_3$ single crystal at 1.8 K. MR is a non-monotonic, with a maximum at 1.18 GPa and 9 T. (**c**) The oscillatory component $\Delta R_{xx}$ at different pressures, extracted from $R_{xx}$ by subtracting a smooth background. (**d**) Fast Fourier transform (FFT) results for SdH oscillations at different pressures. From 0.14 to 0.69 GPa, the $\alpha$ and $\beta$ bands shift slightly to higher frequency, then disappear suddenly at higher pressures, where two higher frequencies $\varepsilon$ and $\xi$ emerge. (**e**) Pressure dependence of FFT frequency. The shaded area contains a Lifshitz transition. (**f**) Landau level index plots for the $\varepsilon$ band at 0.99 GPa, and the $\xi$ band at 1.47 GPa. Both bands have trivial Berry phase.

**Fig. 3 | Pressure-induced superconductivity in SrAs₃.** (**a**) Temperature dependence of the resistance of $SrAs_3$ powder obtained by crushing a self-flux-grown single crystal. (**b**) Magnetic field dependence of the superconducting transition of $SrAs_3$ at 39.4 GPa. (**c**) Contour plot of the pressure dependent of the superconducting transition. (d) Temperature dependence of the upper critical field $\mu_0H_{c2}$. The superconducting transition temperatures ($T_c$s) is defined according to the 10% drop of the transition ($T_c^{10\%}$), the 50% drop of the transition ($T_c^{50\%}$), and zero resistance ($T_c^{zero}$), respectively. The red line is the fit according to the Ginzburg-Landau theory, $\mu_0H_{c2}(T) = \mu_0H_{c2}(0)\ (1-(T/T_c)^2)/(1+(T/T_c)^2)$.

**Fig. 4 | X-ray diffraction (XRD) results of SrAs$_3$ powder under pressure.** (**a**) Selected AD-XRD patterns of SrAs$_3$ at room temperature (λ = 0.6199 Å) up to 51.9 GPa. The ambient-pressure $C2/m$ structure remains stable up to ~ 20.0 GPa. As the pressure further increasing to 23.3 GPa, the appearance of new diffraction peaks (marked with dashed line and asterisk) indicates the emergence of pressure-induced structural phase transition ($C2/m \rightarrow Pm$-$3m$ (SrBi$_3$-like structure)) in SrAs$_3$. (**b**) and (**d**) The schematic crystal structure of ambient-pressure (AP) and high-pressure (HP) phases. The coordination number (CN) of Sr in two phases is 7 and 12, respectively. (**c**) and (**e**) Rietveld refined XRD patterns for 3.6 and 49.8 GPa, respectively. The AP and HP phases coexist at 49.8 GPa. The vertical lines denote the theoretical positions of the Bragg peaks. The different curves between observed and calculated XRD patterns are shown at the bottom.

**Fig. 5 | Band structure for the high-pressure $Pm$-$3m$ phase of SrAs$_3$.** (**a**) The energy band of SrAs$_3$ without turning on SOC effect. Between $\Gamma$ and $M$, the conduction and valence bands cross linearly close to the $E_F$. (**b**) The energy band of SrAs$_3$ with turning on SOC effect. The insets show the detailed band dispersion near the band crossing points. The red dashed line corresponds to a Fermi curve across the gap. (**c**) The bulk BZ and the associated (001) projected BZ with high-symmetry points. The ± symbols represent the party of all occupied bands at each TRIM. (**d**) The topological surface states of SrAs$_3$ on the (001) surface (the green region is enlarged in (**e**)). (**f**) The Fermi surface at - 0.16 eV (marked in (**e**)), the blue arrows are the spin texture.